\documentclass[twocolumn,aps,prb,showpacs]{revtex4-1}
\usepackage{amsfonts}
\usepackage{amsmath}
\usepackage{graphicx}
\usepackage{bm}
\usepackage{amssymb}
\usepackage{times}

\makeatletter

\begin{document}

\title{Thermoelectric Properties of Silicon Carbide Nanowires with
Nitrogen Dopants and Vacancies}
\author{Zhuo Xu}
\author{Qing-Rong Zheng}
\author{Gang Su}
\email[Corresponding author. ]{Email: gsu@gucas.ac.cn}
\affiliation{Theoretical Condensed Matter Physics and Computational
Materials Physics Laboratory, College of Physical Sciences, Graduate
University of Chinese Academy of Sciences, P.O. Box 4588, Beijing
100049, China}

\begin{abstract}
The thermoelectric properties of cubic zincblend silicon carbide
nanowires (SiCNWs) with nitrogen impurities and vacancies along
[111] direction are theoretically studied by means of atomistic
simulations. It is found that the thermoelectric figure of merit ZT
of SiCNWs can be significantly enhanced by doping N impurities
together with making Si vacancies. Aiming at obtaining a large ZT,
we study possible energetically stable configurations, and disclose
that, when N dopants locate at the center, a small number of Si
vacancies at corners are most favored for n-type nanowires, while a
large number of Si vacancies spreading into the flat edge sites are
most favored for p-type nanowires. For the SiCNW with a diameter of
1.1 nm and a length of 4.6 nm, the ZT value for the n-type is shown
capable of reaching 1.78 at 900K. The conditions to get higher ZT
values for longer SiCNWs are also addressed.
\end{abstract}

\maketitle

\section{Introduction}

Recent studies on the thermoelectric cooling and power generation by
semiconductors as solid-state portable energy converters are
prosperous, and there has been a great leap in both theoretical and
experimental techniques. \cite{review1,review2} Practical
applications require that the thermoelectric figure of merit ZT
should be greater than 1.5 (e.g. Ref. \onlinecite{review2}). It is
conceived that one of main approaches to promote ZT is to utilize
low-dimensional semiconductor materials. \cite{2D,1D} For
one-dimensional (1D) quantum wires, as the scattering of confined
electrons is dramatically avoided while the phonons are strongly
scattered off the surface, the ZT could be significantly enhanced,
\cite{1D} which was supported experimentally in doped rough silicon
nanowires (SiNWs), \cite{SiNW} Bi$_{2}$Te$_{3}$ nanowires
\cite{Bi2Te3} and Bi nanowires. \cite{Bi1,Bi2}

On the other hand, silicon carbide (SiC) materials are attracting
much attention for their thermoelectric properties. The nonmetallic
semiconductor SiC shows excellent mechanical properties,
\cite{SiCcomposite} chemical durability, and in particular, the high
temperature stability. Owing to its wide bandgap and low intrinsic
carrier concentration, the semiconducting behavior of SiC can be
kept at temperatures much higher than the case of Si, thus resulting
in a higher operation temperature tolerance for SiC nanodevices.
\cite{SiCNW} Considering also the fact that an effective recovery of
waste heat from vehicle exhaust requires an operating temperature
about 350 $^{\circ}$C, \cite{review2} and other applications such as
high-temperature media-compatible flow sensors, \cite{SiCsensor} one
can see that it is quite necessary to exploit the thermoelectric
properties of low-dimensional SiC materials.

As early as in 2003, Yoshida \emph{et al.} have successfully doped N
and B into 300 $\mu$m thick SiC film that contains Si and C
vacancies, and observed that with N dopants, the power factor of the
SiC film is raised nearly one order of magnitude at 973 K.
\cite{SiCfilm} For SiC nanowires (SiCNWs), the techniques of
fabrication \cite{SiCNW} and measurements \cite{measurement} have
been remarkably improved recently, and theoretical analyses on
structural properties, \cite{structure} mobility, \cite{mobility}
thermal conductivity, \cite{tconductivity} \emph{etc.}, were also
carried out for a few cases. These studies suggest that it is
prospective to obtain nice thermoelectric performance on SiCNWs. To
achieve this goal, the detailed simulations and studies on
thermoelectric properties of SiCNWs are really essential, which is
however sparse in literature.

In the present work, we will focus on the 3C-SiCNWs doped with N
impurities and vacancies along [111] direction with an energetically
favorable hexagonal cross section, \cite{structure} which are
readily fabricated. \cite{3C1,3C2} We investigate the structures,
electronic and phonon transmissions, and analyze the effect of
defects on the thermoelectric performance of SiCNWs. It is found
that the thermoelectric figure of merit ZT of SiCNWs can be
significantly enhanced by doping N impurities together with making
Si vacancies. An optimal doping strategy is suggested for both
n-type and p-type SiCNWs.

This paper is organized as follows. In Sec. II, the simulation
method and details will be described. In Sec. III, the structural
stability of SiCNWs with defects will be discussed. The electronic
structures and transport properties of SiCNWs with various defects
are shown in Sec. IV. In Sec. V, the phonon transport properties of
SiCNWs will be analyzed. In Sec. VI, the effects of different
defects on thermoelectric transport properties of SiCNWs are
presented. Finally, a summary will be given.

\section{Calculational Method}

Structural optimizations are performed by the SIESTA code,
\cite{siesta} which is based on the density functional theory (DFT)
\cite{DFT} with norm-conserving pseudopotentials \cite{T-M} and
linear combinations of atomic orbitals. The optimizations are
spin-polarized, employing double-zeta polarized basis sets within
generalized gradient approximation (GGA) expressed by PBE
functional. \cite{PBE} The energy cutoff is 180 Ry, and the force
tolerance criterion is 0.04 eV/${\mathring{A}}$ for structural
relaxation. The separation between neighboring nanowire surfaces is
15 ${\mathring{A}}$.

The electron transmission spectra are calculated by TRANSIESTA code,
\cite{transiesta} based on the Landauer-B\"{u}ttiker and
nonequilibrium Green's function (NEGF) formalism, \cite{Datta,negf}
employing single-zeta basis sets with GGA and PBE. The energy cutoff
is 100 Ry, and the convergence criterion of density matrix is 0.005.
The leads of the transport model are assumed to be pristine
nanowires. To calculate the mean free path (MFP) $l_{e}$ and
localization length $\xi$ of single N dopant, the supercell between
the leads is chosen to contain five unit cells with a total length
of 38.5 ${\mathring{A}}$. The three middle units together are fully
relaxed, where the single N dopant is in the central unit. Following
the method introduced in Refs. [\onlinecite{MFP1,MFP2}], the average
scattering resistance at different positions for single dopant is
defined as
\begin{equation}
\langle R_{s}(E)\rangle=\sum\limits_{i=1}^{M}\frac{p_{i}}{G_{i}(E)}-R_{c}(E),
\end{equation}%
\begin{equation}
R_{c}(E)=\frac{1}{G_{0}(E)}=\frac{h}{2e^{2}N(E)},
\end{equation}%
where $R_{s}$, $R_{c}$, $G_{i}$, $G_{0}$, $p_{i}$ and $N$ are
scattering resistance, contact resistance, conductance with dopants,
conductance without dopant, the weight of different doping
positions, and the number of conducting channels, respectively. The
mean resistance of wire and thus the MFP can be estimated linearly
by
\begin{equation}
l_{e}(E)=\frac{R_{c}(E)}{\langle R_{s}(E)\rangle}d,
\end{equation}%
where $d$ is the average dopant-dopant separation estimated from a
realistic doping density. Such a linear relation is valid when the
wire length $L$ is in the quasiballistic ($L<l_{e}$) and diffusive
($l_{e}<L<\xi$) regimes. In the localization regime ($L>\xi$), the
resistance increases exponentially. \cite{MFP1,MFP2} For 3C-SiC, the
carrier concentration can be controlled in the range of
10$^{15}$-10$^{19}$ cm$^{-3}$ (Ref. [\onlinecite{concentration}]).
In the present work we use $d$=40 nm corresponding to a bulk doping
density about 1.5$\times$10$^{16}$ cm$^{-3}$.
The localization length is calculated by
\cite{xi}
\begin{equation}
\xi(E)=\frac{1}{2}[N(E)+1]l_{e}(E).
\end{equation}

To calculate the transport properties of the SiCNWs with different
defect combinations of N dopants and vacancies, we suppose that the
defects are distributed homogeneously in the units between the
leads, where the length between the leads is taken as 3 units (about
2.3 nm) by default, except for the case of 6 units with particular
specification. For phonon transmission, the force constant matrices
are calculated by GULP code, \cite{gulp} which is based on the
Tersoff model of empirical potential (TEP).
\cite{tersoff1,tersoff2,tersoff3} To obtain dynamical matrices, the
intra-atomic elements of the force constant matrices output from
GULP are replaced by the data recalculated under the condition of
momentum conservation, \cite{SiNWt} where the atomic masses are then
included. The precision criterion of the matrices is 10$^{-5}$. In
terms of the dynamical matrices, the phonon transmission spectra are
calculated within the Landauer-B\"{u}ttiker and NEGF formalism
similar to the case of electrons, following the method described in
Refs. [\onlinecite{SiNWt,ttransport,sgf}]. It is known that the
Landauer-B\"{u}ttiker formalism employs a finite central region for
transport. In our calculations we use the nanowires with a fixed
length to get the electronic and thermal conductance from the
electron and phonon transmission spectra, where the length can be
eliminated in calculations of the thermoelectric figure of merit ZT
that can be obtained by \cite{review1,SiNWte,BiNW}
\begin{equation}
ZT=\frac{S^{2}\sigma_{e}T}{\kappa_{ph}+\kappa_{e}},
\end{equation}%
\begin{equation}
\kappa_{ph}=\frac{\hbar^{2}}{2\pi k_{B}T^{2}}\int_{0}^{\infty}d\omega\omega^{2}
\mathcal{T}(\omega)\frac{e^{\hbar\omega/k_{B}T}}{(e^{\hbar\omega/k_{B}T}-1)^{2}},
\end{equation}%
\begin{equation}
\sigma_{e}(\mu)=\frac{2e^{2}}{hk_{B}T}\int_{-\infty}^{\infty}dE
\mathcal{T}(E)\frac{e^{(E-\mu)/k_{B}T}}{[e^{(E-\mu)/k_{B}T}+1]^{2}}=e^{2}\mathcal{L}^{(0)},
\end{equation}%
\begin{equation}
\mathcal{L}^{(\alpha)}(\mu)=\frac{2}{hk_{B}T}\int_{-\infty}^{\infty}dE
\mathcal{T}(E)(E-\mu)^{\alpha}\frac{e^{(E-\mu)/k_{B}T}}{[e^{(E-\mu)/k_{B}T}+1]^{2}},
\end{equation}%
\begin{equation}
S(\mu)=\frac{\mathcal{L}^{(1)}(\mu)}{eT\mathcal{L}^{(0)}(\mu)},
\end{equation}%
\begin{equation}
\kappa_{e}(\mu)=\frac{1}{T}\{\mathcal{L}^{(2)}(\mu)-\frac{[\mathcal{L}^{(1)}(\mu)]^{2}}
{\mathcal{L}^{(0)}(\mu)}\},
\end{equation}%
where $S$ is the Seebeck coefficient, $T$ is the temperature, $\mu$ is
the chemical potential, $\sigma_{e}$, $\kappa_{e}$ and $\kappa_{ph}$
are the electronic conductance, thermal conductance of electrons and
phonons, respectively, $\mathcal{T}(\omega)$ and $\mathcal{T}(E)$
are the phonon and electron transmission, respectively.
$S^{2}\sigma_{e}$ is the power factor. Generally
speaking, the closer of the chemical potential $\mu$ to the
conduction band minimum (CBM) or the valence band maximum (VBM) is,
the higher carrier concentration of electrons or holes is.
\cite{SiNWte,GeSiNWte}
The nonlinear effects such as electron-electron, phonon-phonon and
electron-phonon interactions are ignored for simplicity in the
present work. \cite{ttransport,SiNWte}

\section{Structural Stability with Defects}

In experimental and theoretical studies on SiCNWs, the most reported
one is the cubic zincblend 3C($\beta$)-SiCNWs along [111] direction,
because it can be deposited on Si.
\cite{SiCNW,SiCNWfet,structure,modulus2} As a typical model for
simulation, we choose the ultrathin 3C-SiCNWs with a diameter of 1.1
nm and a hexagonal cross section to investigate their structural and
thermoelectric properties. As shown in Fig. 1(a), one unit cell of
the pristine SiCNW contains 37 Si and 37 C atoms in pairs along the
[111] direction, and 42 H atoms are covered on surface to saturate
the dangling bonds of Si and C atoms. Following from the simulations
on SiNWs, \cite{SiNWt,SiNWte,SiNWH} it is necessary to include H
atoms as surface passivation for electronic calculations, while the
H atoms could be omitted for phonon calculations that produces a
small deviation no larger than 3\%. \cite{SiNWt,SiNWte}

\begin{figure}
\includegraphics[width=0.7\linewidth,clip]{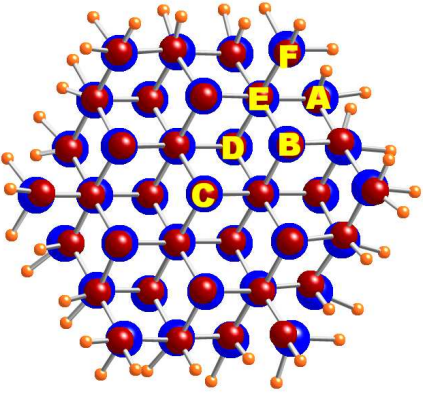}
\caption{(Color online) The cross section of pure SiCNWH along
the axial direction [111], with the marks of different doping sites.
The blue,
dark red, orange and light green balls represent Si, C, H and N
atoms, respectively.} \label{1}
\end{figure}

The properties of doping one N or B atom per unit cell into the
3C-SiCNW have been theoretically investigated in Ref.
[\onlinecite{SiCNWdoping}]. It reveals that it is more favorable to
dope N atom substituting C than substituting Si atom, the
energetically most stable doping site of N is at the center of the
SiCNW (site C), and the inner doping positions of B, C and D are
energetically more favorable than the near-surface position E and
surface positions A and F, \cite{SiCNWdoping} which is in agreement
with our results.

We go on investigating the combined defects with N impurities and
vacancies in the SiCNW. Such kinds of defects have been successfully
implanted in SiC film, \cite{SiCfilm} but no details on the
distribution of doping positions were given. Here, the sites of
impurities in the SiCNW are marked in Fig. 1. For clarity, the
case that dopes one N atom per unit cell substituting the C atom at
the center (site C) is labeled as SiCNWN$_{C}$H; the case that dopes
two N atoms substituting C atoms at sites C and A is labeled as
SiCNWN$_{CA}$H; the case that dopes one N at site C and one Si
vacancy at site A with saturated passivation by H atoms is labeled
as SiCNWN$_{C}$V$^{Si}_{A}$H, where the case
with unsaturated passivation leaving 3 dangling bonds per unit cell
is labeled as SiCNWN$_{C}$V$^{Si}_{A}$h(3), and leaving 1 dangling
bond as SiCNWN$_{C}$V$^{Si}_{A}$h(1); the case that dopes one N at C
site and three Si vacancies at three corresponding A sites with
saturated passivation is labeled as SiCNWN$_{C}$V$^{Si}_{3A}$H; and
so on. When vacancies are included, a supercell containing two units
is employed where neighboring units have the interlaced vacancies at
the corresponding symmetrical sites. Note that one dangling bond per
unit cell in SiCNWN$_{CA}$H is not left by the SiC, but by the N
dopant at the surface site A.

The formation energy E$^{f}$ averaged over one unit cell is
calculated by \cite{SiCNWdoping,formation1,formation2}
\begin{equation}
E^{f}(D)=E^{tot}(D)-E^{tot}(pure)-\Sigma _{i}\Delta n_{i}\mu_{i},
\end{equation}%
where E$^{tot}$(D) and E$^{tot}$(pure) represent the total energy
with and without the defects (D), respectively, $\Delta n_{i}$ is
the number increment of atoms induced by the defects, $\mu _{i}$ is
the chemical potential, and $i=N,C,Si$. At Si-rich, N-rich and
H-rich limit, the chemical potentials are $\mu _{Si}=\mu
_{Si}^{bulk}$, $\mu _{C}=\mu _{SiC}^{bulk}-\mu _{Si}$, $\mu _{N}=\mu
_{N_{2}}/2$ and $\mu _{H}=\mu _{H_{2}}/2$, respectively. The
formation energies of several typical cases are shown in Table I.

\begin{table}[tb]
\caption{The formation energies E$^{f}$ of SiCNWs with defects of
N and vacancy under Si-, N- and H-rich conditions.}
\begin{tabular}{ccccc}
\hline\hline
NW with defect & E$^{f}$(eV) & NW with defect & E$^{f}$(eV) \\
\hline
SiCNWN$_{C}$H & -0.75 & SiCNWN$_{C}$V$^{C}_{A}$H & 1.86 \\
SiCNWN$_{CA}$H & -1.38 & SiCNWN$_{C}$V$^{Si}_{F}$H & -1.37 \\
SiCNWN$_{CD}$H & -1.36 & SiCNWN$_{C}$V$^{Si}_{3F}$H & -2.51 \\
SiCNWN$_{C}$V$^{Si}_{A}$h(1) & -0.01 & SiCNWN$_{C}$V$^{Si}_{A}$H & -1.26 \\
SiCNWN$_{C}$V$^{Si}_{A}$h(3) & 4.33 & SiCNWN$_{C}$V$^{Si}_{3A}$H & -1.45 \\
SiCNWN$_{C}$V$^{Si}_{C}$h & 3.89 & SiCNWN$_{A}$V$^{Si}_{A}$H & -0.86 \\
\hline\hline
\end{tabular}
\end{table}

One may see that SiCNWN$_{C}$V$^{Si}_{A}$h(3),
SiCNWN$_{C}$V$^{Si}_{C}$h and SiCNWN$_{C}$V$^{C}_{A}$H have very
large positive values of E$^{f}$, implying that these structures
with defects are unlikely to form because of too much formation
energy needed. \cite{formation1,formation2}
SiCNWN$_{C}$V$^{Si}_{A}$h(1) has an E$^{f}$ very close to zero,
showing that it is also unstable. All of the remaining cases have
negative E$^{f}$, which are energetically favorable and would be
likely to be fabricated. The common ground of these cases is that
all the Si and C atoms get saturated passivation. It appears that
the Si vacancies favor the surface sites where it is easy to
passivate all the dangling bonds, thus enabling those structures
stable. Comparing SiCNWN$_{A}$V$^{Si}_{A}$H with
SiCNWN$_{C}$V$^{Si}_{A}$H, in presence of the Si vacancy on the
surface, the N dopant is obviously more favorable for the center
site than the surface. Comparing SiCNWN$_{CA}$H and SiCNWN$_{CD}$H,
SiCNWN$_{C}$V$^{Si}_{F}$H and SiCNWN$_{C}$V$^{Si}_{A}$H, we note
that in presence of one N dopant stable at the center, the E$^{f}$
differences between them are as small as 0.02 eV and 0.11 eV,
respectively, suggesting that their corresponding concentrations in
the SiCNWs are close. However, E$^{f}$ of SiCNWN$_{C}$V$^{Si}_{3F}$H
is much lower than that of SiCNWN$_{C}$V$^{Si}_{3A}$H by 1.06 eV. It
is clear that the location of Si vacancies prefers energetically the
corner (site F) to the flat edge (site A).

\begin{figure}
\includegraphics[width=1.0\linewidth,clip]{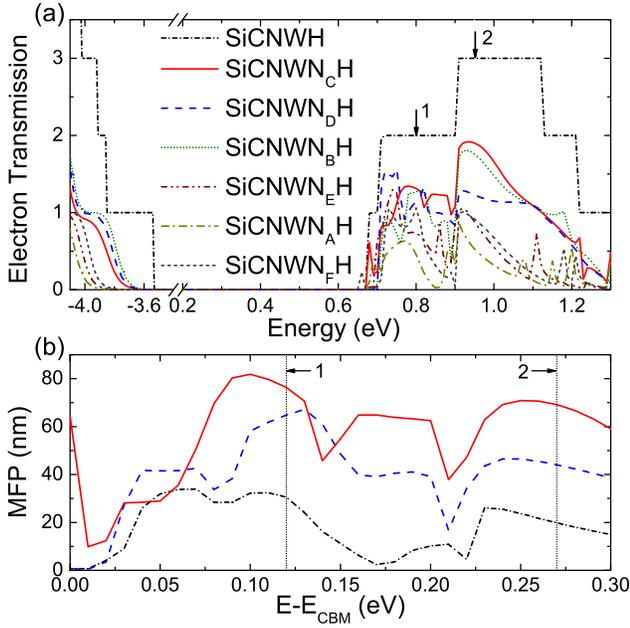}
\caption{(Color online) (a) The electronic transmission spectra of
SiCNWs with one single N dopant in the center of the supercell of 5
units between the leads, doped at positions A, B, C, D, E and F,
represented by dark yellow dash-dotted, green dotted, solid red,
blue dashed, wine dash-dot-dotted and dark gray shot dashed curves,
respectively. The black short dash-dotted curve represents the
defect-free SiCNWH. (b) The MFP of single N dopant of three cases:
(i) the average of all six doping positions (black short dash-dotted
curve), (ii) the average of the positions B, C and D (blue dashed
curve), and (iii) the position C (red solid curve), where $d$=40 nm.
In both (a) and (b), the energy values of E=0.8 and 0.95 eV are
marked as 1 and 2, respectively.} \label{1}
\end{figure}

To estimate the MFP $l_{e}$ and the localization length $\xi$ for
the case with single N dopant substituting one C atom, the
electronic transmission spectra of the cases with one single N atom
doped at different positions are calculated, as shown in Fig. 2(a).
Except for the spectra very close to the CBM,
the spectra associated with the three inner doping positions
B, C and D are obviously higher than those of the surface and
near-surface positions A, E and F. In Fig. 2(b), the MFPs for the
structures with single N dopant of (i) the average of all six doping
positions, (ii) the average of the positions B, C and D, and (iii)
the position C are presented, where a dopant-dopant separation $d$
is taken as 40 nm. The energetically stability of these three cases
strengthens from (i) to (iii) \cite{SiCNWdoping}. From Eqs. (3) and
(4), we find that at E-E$_{CBM}$=0.12 with the conducting channels
N=2, for case (i), (ii) and (iii), $l_{e}$=30.58, 64.78 and 76.35
nm, $\xi$=45.87, 97.17 and 114.53 nm, respectively; and at
E-E$_{CBM}$=0.27 with N=3, $l_{e}$=19.84, 44.05 and 69.22 nm,
$\xi$=39.69, 88.10 and 138.44 nm, respectively. It indicates that
for single N dopant, the structures with energetically favorable
inner doping positions have the MFPs and localization lengths larger
than those with the surface positions that are energetically
unfavorable.

\section{Electronic Structure and Transport}

\begin{figure}
\includegraphics[width=1.0\linewidth,clip]{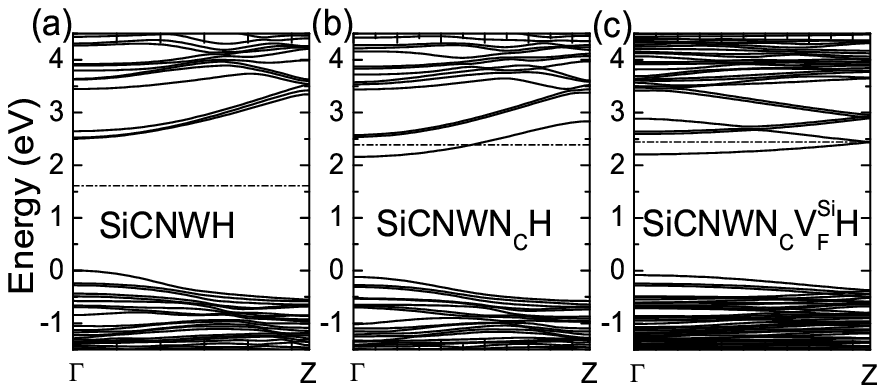}
\caption{(Color online) The electronic band structure of (a) SiCNWH,
(b) SiCNWN$_{C}$H and (c) SiCNWN$_{C}$V$^{Si}_{F}$H (a supercell of
2 units). The zero point of energy is set to the VBM of SiCNWH. The
corresponding Fermi levels are marked by dash-dotted lines.}
\label{1}
\includegraphics[width=1.0\linewidth,clip]{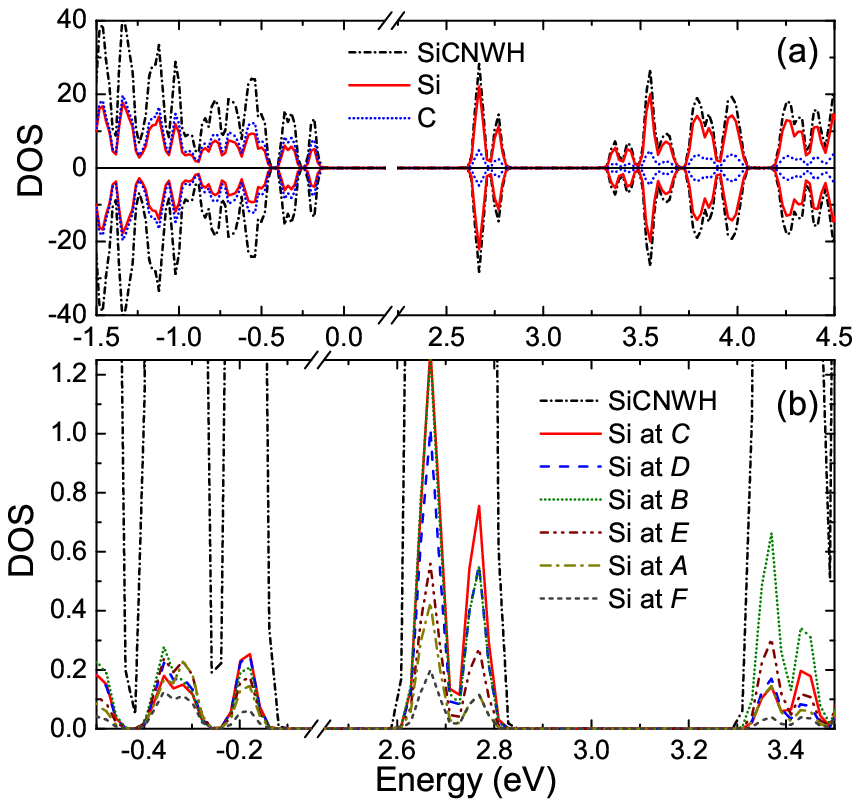}
\caption{(Color online) (a) The spin-polarized DOS of SiCNWH (black short
dash-dotted curve) and the PDOS of Si (red solid curve) and C (blue dotted curve)
atoms. (b) The spin-up DOS of SiCNWH, and the PDOS of one Si atom at
positions A, B, C, D, E and F, respectively. The curve marks are in accordance
with Fig. 2(a). The zero point of energy is set to the VBM of SiCNWH.}
\label{1}
\end{figure}

The electronic band structures of SiCNWH, SiCNWN$_{C}$H and
SiCNWN$_{C}$V$^{Si}_{F}$H are shown in Fig. 3, where the zero point
of energy is set to the VBM of the defect-free SiCNWH. From Fig.
3(a), one may see that SiCNWH has a direct gap, where the Fermi
level locates in the band-gap. Fig. 3(b) manifests that the N dopant
in SiCNW is an n-type defect, which significantly changes the bottom
of the conduction band, and the Fermi level is cross this CBM.
\cite{SiCNWdoping} Therefore, the N impurity gives an essential
contribution to the electronic transport. It is noted that the Fermi
level of SiCNWN$_{C}$H is 0.12 eV lower than the CBM of the
defect-free SiCNWH, and the degeneracy of the bands of SiCNWN$_{C}$H
is also lower than that of SiCNWH. Similar changes occur for the
case of SiCNWN$_{C}$V$^{Si}_{F}$H shown in Fig. 3(c). The n-type
defects introduce more electron carriers and stronger scattering of
electrons.

Fig. 4(a) shows the density of states (DOS) of the defect-free
SiCNWH, which is calculated with a spin-polarized code. It can be
seen that the electronic band is spin unpolarized, and the Si atoms
contribute overwhelmingly to the DOS at the bottom of the conduction
band. The peaks of the projected DOS (PDOS) of Si and C atoms
totally overlap, and their relative magnitudes are inverse in the
valence and conduction bands, indicating that the Si and C atoms
bond with each other. The PDOS of one Si atom at different positions
is presented in Fig. 4(b). For the two peaks of DOS around 2.7 eV,
the PDOS of the system with one Si atom at position B is almost
coincident with that of Si at position C in the lower peak and with
that of Si at position D in the higher peak. The PDOS of Si at
positions A and F in the higher peak are also nearly coincident. The
PDOS of Si at inner positions B, C and D are significantly higher
than those at outer positions E, F and A.

It is interesting to note that for the peak of electronic
transmission spectra from 0.8 to 1.2 eV in Fig. 2(a), the spectra of
the system with N dopant substituting C atom at inner positions B, C
and D are obviously higher than those at outer positions E, F and A,
which is qualitatively consistent with the PDOS of different Si
atoms for the peak around 2.7 eV in Fig. 4(b). When the N dopant
substitutes one C atom in SiCNWs, the neighboring Si atoms connected
by bonds will obtain extra electrons from the dopants. The PDOS of
the N dopant will spread to the valence and conduction bands similar
to the PDOS of Si atoms. \cite{SiCNWdoping} At the bottom of the
conduction band, the Si atoms at inner positions contribute to the
majority of the DOS, so the N defects at inner positions contribute
electrons to the bottom of conduction bands more than those doped at
outer positions. From these results one may deduce that for the peak
of electronic transmission spectra at the bottom of conduction band,
the spectra with N defect at inner positions could be higher than
those at outer positions, which is just the results shown in Fig.
2(a).

\begin{figure}
\includegraphics[width=1.0\linewidth,clip]{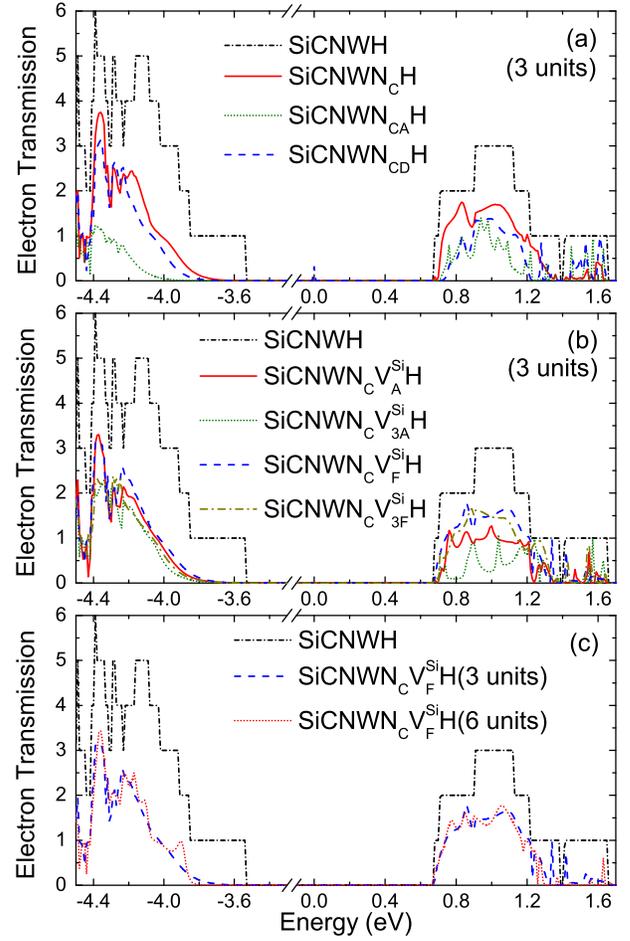}
\caption{(Color online) The electronic transmission spectra of
SiCNWs with various defects, where the black short dash-dotted curve
always represent pure SiCNWH. (a) SiCNWN$_{C}$H, SiCNWN$_{CA}$H and
SiCNWN$_{CD}$H are represented by red solid, green dotted and blue
dashed curves, respectively; (b) SiCNWN$_{C}$V$^{Si}_{A}$H,
SiCNWN$_{C}$V$^{Si}_{3A}$H, SiCNWN$_{C}$V$^{Si}_{F}$H and
SiCNWN$_{C}$V$^{Si}_{3F}$H are represented by red solid, green
dotted, blue dashed and dark yellow dash-dotted curves,
respectively. (c) The electronic transmission spectra of
SiCNWN$_{C}$V$^{Si}_{F}$H of 3 units (blue dashed curve) and 6 units
(red dotted curve) for a comparison.} \label{1}
\end{figure}

To probe the transport properties of SiCNWs with defects, besides
the pure SiCNWH and SiCNWN$_{C}$H that is energetically the most
stable case for the N$_{C}$ defect, \cite{SiCNWdoping}
SiCNWN$_{CA}$H, SiCNWN$_{CD}$H, SiCNWN$_{C}$V$^{Si}_{A}$H,
SiCNWN$_{C}$V$^{Si}_{F}$H, SiCNWN$_{C}$V$^{Si}_{3A}$H and
SiCNWN$_{C}$V$^{Si}_{3F}$H are selected as relatively stable
examples from the cases with different doping sites.

The electronic transmissions of SiCNWs with various defects are
given in Fig. 5. As shown in Fig. 5(a), for SiCNWN$_{C}$H, at the
bottom of conduction band the electron transmission spectrum shrinks
slightly from the spectrum of SiCNWH, while at the top of valence
band the hole transmission spectrum has a significant shrink from
that of SiCNWH. For SiCNWN$_{CA}$H and SiCNWN$_{CD}$H the
transmission spectra go on shrinking from SiCNWN$_{C}$H, and the
shrink is extraordinarily evident for the hole transmission of
SiCNWN$_{CA}$H that has one surface dangling bond left by the N
dopant at site A. So the N impurities in the SiCNW induce stronger
scattering of electrons, which overcomes the effect of the increase
of electron carriers by the n-type doping, resulting in the decrease
of electron/hole transmission. More dopants of N will lead to a
greater decrease of transmission.

The cases of SiCNWs with implantation of saturated Si vacancies on
surface in addition to the N dopant at center are shown in Fig.
5(b). The spectra of SiCNWN$_{C}$V$^{Si}_{3A}$H and
SiCNWN$_{C}$V$^{Si}_{3F}$H with more Si vacancies are a bit lower
than those of SiCNWN$_{C}$V$^{Si}_{A}$H and
SiCNWN$_{C}$V$^{Si}_{F}$H owing to the stronger scattering by the
vacancies. Comparing the spectra of SiCNWN$_{C}$V$^{Si}_{A}$H,
SiCNWN$_{C}$V$^{Si}_{3A}$H, SiCNWN$_{C}$V$^{Si}_{F}$H and
SiCNWN$_{C}$V$^{Si}_{3F}$H in Fig. 5(b) with that of SiCNWN$_{C}$H
in Fig. 5(a), we find that in presence of the central N dopant, for
the hole transmission with surface Si vacancies at either the corner
(site F) or the flat edge (site A) site, and for the electronic
transmission with surface Si vacancies at the corner F, the spectra
change little from that of SiCNWN$_{C}$H, revealing that the
electron or hole scattering are not significantly strengthened by
the vacancies. Consequently, the MFP and the localization length of
SiCNWN$_{C}$V$^{Si}_{F}$H and SiCNWN$_{C}$V$^{Si}_{3F}$H would have
values close to those of SiCNWN$_{C}$H due to their similar
transmission spectra.

Fig. 5(c) compares the electronic transmission spectra of
SiCNWN$_{C}$V$^{Si}_{F}$H with 3 and 6 units in length. The two
spectra are quite close, and the spectrum of 6 units has a larger
gradient at the edge of the bands. It reveals that for
SiCNWN$_{C}$V$^{Si}_{F}$H, the effects of increasing electron
carriers and strengthening the electron scattering by the n-type
defects get balanced so that the electronic transmission changes
little, as the wire length increases.

\begin{figure}
\includegraphics[width=1.0\linewidth,clip]{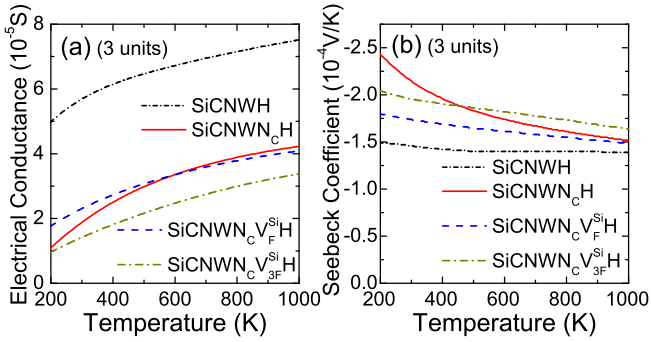}
\caption{(Color online) The temperature dependence of (a) the
electrical conductance $\sigma_{e}$ and (b) the Seebeck coefficient
$S$ for pure SiCNWH (black short dash-dotted curve),
SiCNWN$_{C}$H (red solid curve), SiCNWN$_{C}$V$^{Si}_{F}$H (blue
dashed curve) and SiCNWN$_{C}$V$^{Si}_{3F}$H (dark yellow
dash-dotted curve), respectively. The chemical potential $\mu$ is
set to 0.675 eV, which is the bottom of the electron transmission
spectrum of pure SiCNWH as indicated in Figs. 5(a) and (b).}
\label{1}
\end{figure}

With the electronic transmission spectra, the electrical conductance
$\sigma_{e}$, the Seebeck coefficient $S$ and the power factor
$S^{2}\sigma_{e}$ can be calculated by means of Eqs. (7) and (9).
Generally, the maximum of ZT is achieved when the chemical potential
$\mu$ is around the CBM for electrons and the VBM for holes
\cite{SiNWte}. The temperature dependences of $\sigma_{e}$ and
\textbf{$S$} are presented in Figs. 6(a) and (b) for pure SiCNWH,
SiCNWN$_{C}$H, SiCNWN$_{C}$V$^{Si}_{F}$H and
SiCNWN$_{C}$V$^{Si}_{3F}$H, respectively. These four cases are
chosen owing to their relatively large transmission of electrons as
seen from Figs. 5(a) and (b). Here $\mu$ is set to 0.675 eV, which
is the bottom of the electron transmission spectrum of pure SiCNWH.
As shown in Figs. 6(a) and (b), the differences among SiCNWN$_{C}$H,
SiCNWN$_{C}$V$^{Si}_{F}$H and SiCNWN$_{C}$V$^{Si}_{3F}$H are small,
unveiling that the effect of the increase of electron carriers and
the enhancement of electron scattering induced by more Si vacancies
are nearly balanced. No matter the defects that include Si vacancies
or not, $\sigma_{e}$ are significantly lower than those of pure
SiCNWH due to the electron scattering by the N dopants. Moreover, it
can be seen that for all four cases, as temperature increases from
200 to 1000 K, the electrical conductance $\sigma_{e}$ increases
while the magnitude of negative Seebeck coefficient $S$ decreases.
When temperature is higher than 400 K, the changes of both
$\sigma_{e}$ and $S$ become slow.

\section{Phonon Transport}

\begin{figure}
\includegraphics[width=1.0\linewidth,clip]{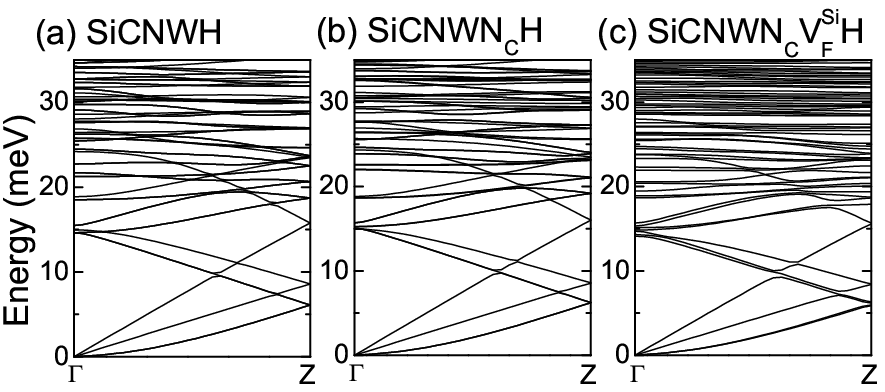}
\caption{(Color online) The phonon band structure of (a) SiCNWH,
(b) SiCNWN$_{C}$H and (c) SiCNWN$_{C}$V$^{Si}_{F}$H, all of which are
calculated in a supercell of 2 units.}
\label{1}
\includegraphics[width=1.0\linewidth,clip]{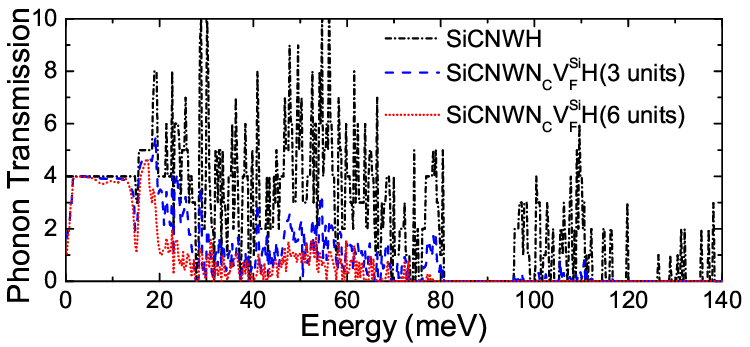}
\caption{(Color online) The phonon transmission spectra of
pure SiCNWH (black short dash-dotted curve), SiCNWN$_{C}$V$^{Si}_{F}$H
with 3 units (blue dashed curve) and 6 units (red dotted curve),
respectively.}
\label{1}
\includegraphics[width=1.0\linewidth,clip]{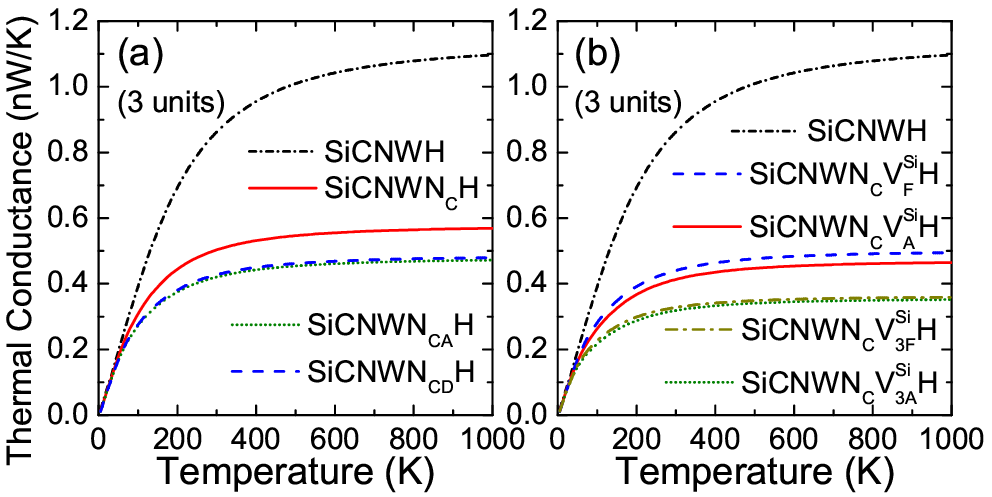}
\caption{(Color online) The lattice thermal conductance $\kappa_{ph}$
of SiCNWs with various defects for temperature from 0 to 1000 K.
The symbols of different curves are totally the same as Figs. 5(a)
and (b).}
\label{1}
\end{figure}

To investigate the properties of phonon transport, Fig. 7 gives the
phonon band structures of SiCNWH, SiCNWN$_{C}$H and
SiCNWN$_{C}$V$^{Si}_{F}$H. It is uncovered that the degeneracy of
phonon bands is lowered either by doping N at the center into
SiCNWH, or by doping Si vacancies at the corner site F into
SiCNWN$_{C}$H, suggesting that the phonon scattering is
strengthened. However, the number and the energy ranges of the
phonon bands are nearly unchanged among these cases, which is in
contrast to the defect bands in the gap of electronic band
structures (Fig. 3).

As shown in Fig. 8, the phonon transmission of pure SiCNWH is equal
to 4 at the low energy range close to zero, which corresponds to the
four acoustic modes of phonons in nanowires. Compared with SiCNWH,
the phonon transmission spectrum of SiCNWN$_{C}$V$^{Si}_{F}$H is
obviously lower, and above 80 meV, the phonon transmission decreases
near to zero. When the wire length increases from 3 to 6 units, the
phonon transmission of SiCNWN$_{C}$V$^{Si}_{F}$H is further lowered.
In contrast to the results in Fig. 5(c) that the electronic
transmission of SiCNWN$_{C}$V$^{Si}_{F}$H change little from 3 to 6
units in length, there are no factors to balance the strengthening
of phonon scattering by the defects, and the phonon transmission of
SiCNWN$_{C}$V$^{Si}_{F}$H decreases significantly as the wire length
increases.

Fig. 9 presents the lattice thermal conductance $\kappa_{ph}$ for
various cases in the temperature range of 0-1000 K. $\kappa_{ph}$
first goes up rapidly and then increases slowly when temperature is
higher than 300 K. It is seen that the $\kappa_{ph}$ decreases as
the defects of N or Si vacancy increase. The differences of
$\kappa_{ph}$ between SiCNWN$_{CA}$H and SiCNWN$_{CD}$H,
SiCNWN$_{C}$V$^{Si}_{A}$H and SiCNWN$_{C}$V$^{Si}_{F}$H,
SiCNWN$_{C}$V$^{Si}_{3A}$H and SiCNWN$_{C}$V$^{Si}_{3F}$H are small,
which are 0.01 nW/K, 0.03 nW/K and 0.01 nW/K at 1000 K,
respectively. It indicates that the same doping concentration at
different doping sites leads to very similar thermal conductance.
Comparing Fig. 6 with the corresponding cases in Fig. 9, one may see
that in the presence of N dopants or Si vacancies, both $\sigma_{e}$
and $\kappa_{ph}$ will decrease from the values of pure SiCNWH, but
the amplitude of $\kappa_{ph}$ is decreased more remarkably than
that of $\sigma_{e}$.

\section{Thermoelectric Transport}

\begin{figure}
\includegraphics[width=1.0\linewidth,clip]{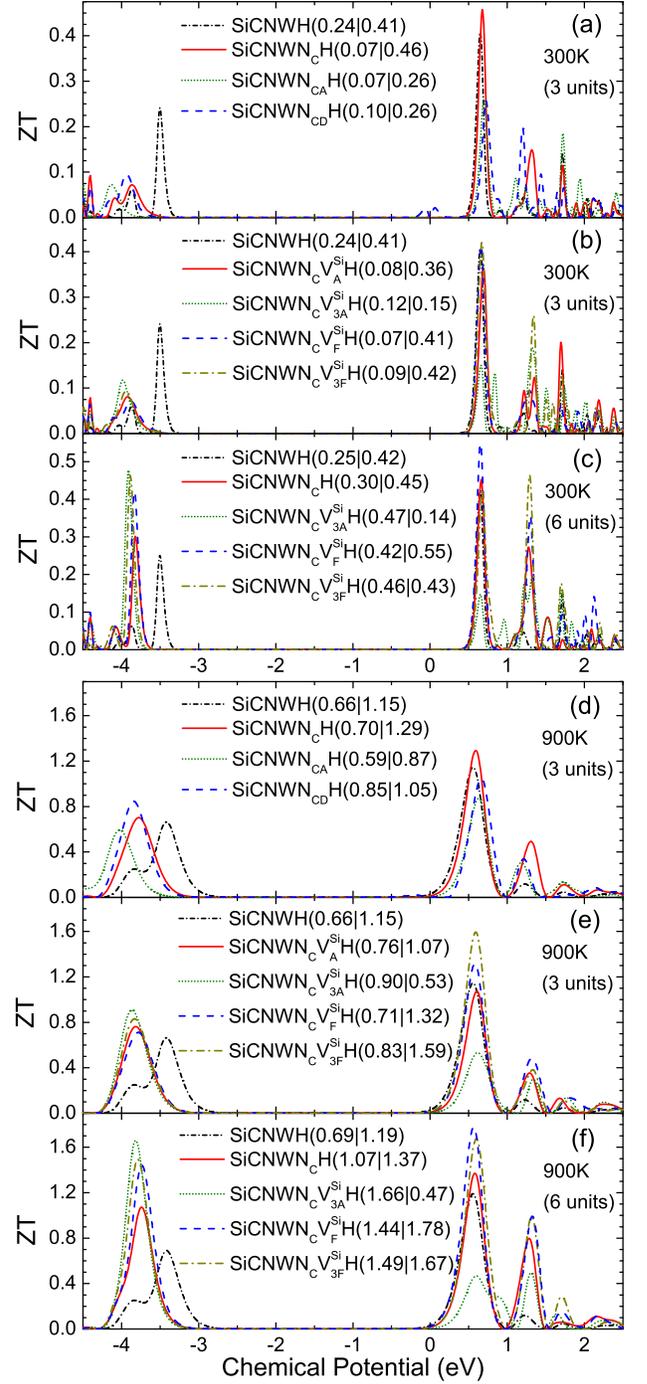}
\caption{(Color online) The thermoelectric figure of merit ZT of
SiCNWs with various defects as a function of chemical potential
$\mu$ at (a)-(c) 300 K and (d)-(f) 900 K. The length of SiCNWs in
(c) and (f) is 4.6 nm of 6 unit cells, and the length of other cases
is 2.3 nm of 3 unit cells. The symbols of different curves are
totally the same as in Figs. 5(a) and 5(b). On both sides of the
zero point of chemical potential, the ZT value of the first peak of
each curve is listed in the bracket after the figure legend of the
corresponding curve. The left (right) value in the bracket
corresponds to the first peak of p-type (n-type) on the left (right)
side.} \label{1}
\end{figure}

Combining the results of electron and phonon transport, the
thermoelectric figure of merit ZT is calculated as a function of the
chemical potential $\mu$ \cite{SiNWte} by Eqs. (5)-(10) at 300 K and
900 K, respectively, as shown in Fig. 10. Comparing Figs. 10 and 5,
the maximum of ZT occurs when the chemical potential reaches the VBM
for p-type SiCNWs or the CBM for n-type SiCNWs, where the electronic
transmission changes dramatically.

For n-type SiCNWs, at 300K or 900K, it is clear that in Figs. 10(a),
10(b), 10(d) and 10(e), the cases of SiCNWN$_{C}$H,
SiCNWN$_{C}$V$^{Si}_{F}$H and SiCNWN$_{C}$V$^{Si}_{3F}$H can induce
obvious enhancements of the ZT. For other cases the ZT is
even lower than that of SiCNWH. Among SiCNWN$_{C}$H,
SiCNWN$_{C}$V$^{Si}_{F}$H and SiCNWN$_{C}$V$^{Si}_{3F}$H, their
relative magnitudes of ZT change as the length increases from 3 to 6
units, as shown in Figs. 10(c) and 10(f). For a length of 6 units, the
n-type ZT maximum of SiCNWN$_{C}$H, SiCNWN$_{C}$V$^{Si}_{F}$H and
SiCNWN$_{C}$V$^{Si}_{3F}$H are 0.45, 0.55 and 0.43 at 300 K, 1.37,
1.78 and 1.67 at 900 K, respectively. Thus, the ZT maximum for
n-type is given by SiCNWN$_{C}$V$^{Si}_{F}$H, containing the defects
of one N dopant at the center and one Si vacancy at the corner per
unit cell. Neither more N dopants such as SiCNWN$_{CD}$H nor more Si
vacancies such as SiCNWN$_{C}$V$^{Si}_{3F}$H are beneficial.
In general, to avoid the ZT dropping from the maximum of
SiCNWN$_{C}$V$^{Si}_{F}$H to the value of SiCNWN$_{C}$V$^{Si}_{3F}$H
or others, it is preferable to dope N impurities limited at the
center and a small quantity of Si vacancies limited at the corners
(F sites), which is energetically favorable. Such a distribution of
Si vacancies at the corners strengthens the phonon scattering
dramatically while it influences little on the electron scattering.

For SiCNWN$_{C}$H and SiCNWN$_{C}$V$^{Si}_{F}$H, it was shown in
Fig. 3 that their Fermi levels are about 0.1 eV lower than the CBM
of the defect-free SiCNWH. The Fermi level equals to the chemical
potential at zero temperature. As manifested in Fig. 10, because the
highest peaks of ZT for n-type accumulate around the CBM of SiCNWH,
the corresponding chemical potential at finite temperature will lead
to a ZT value very close to the maximum of ZT for n-type.

For p-type SiCNWs, among the cases in Figs. 10(a), 10(b), 10(d) and
10(e), the maximum of ZT of SiCNWN$_{CD}$H is larger than that of
SiCNWN$_{CA}$H. Considering the tiny difference of formation energy
E$^{f}$ between SiCNWN$_{CA}$H and SiCNWN$_{CD}$H, the larger ZT of
SiCNWN$_{CD}$H is hard to maintain. On the other hand, the
differences of ZT among SiCNWN$_{C}$V$^{Si}_{A}$H,
SiCNW$N_{C}$V$^{Si}_{3A}$H, SiCNWN$_{C}$V$^{Si}_{F}$H and
SiCNWN$_{C}$V$^{Si}_{3F}$H are quite small, while the energetically
most favorable one is SiCNWN$_{C}$V$^{Si}_{3F}$H. All these four
cases have significant ZT enhancements in Figs. 10(c) and 10(f),
and relatively, the cases with more Si vacancies have larger ZT.
For the best two cases, SiCNWN$_{C}$V$^{Si}_{3A}$H and
SiCNWN$_{C}$V$^{Si}_{3F}$H with a length of 6 units, the p-type ZT
maximum is 0.47 and 0.46 at 300 K, 1.66 and 1.49 at 900 K,
respectively, which are obviously larger than the values of SiCNWH,
as shown in Figs. 10(c) and 10(f). Therefore, it is preferable to
dope N impurities stable at the center and a large quantity of Si
vacancies not only at the corners (F sites) but also spreading to
the flat edges (A sites). However, the p-type performance depends
on the ingredient of the p-type dopants which is not addressed in
the present work.

It should be noted that the selected cases cannot cover all
positions and concentrations of the defects. However, the
performances affected by various defects are reflected in the ZT
curves of these selected interesting cases as functions of $\mu$.
For instance, as displayed in Figs. 10(a) and 10(d), for the n-type
the peaks of ZT for SiCNWN$_{C}$H, SiCNWN$_{CA}$H and SiCNWN$_{CD}$H
appear at $\mu=$ 0.675, 0.695 and 0.725 eV at 300 K, 0.595, 0.635
and 0.665 eV at 900 K, respectively. Compared with SiCNWN$_{C}$H,
SiCNWN$_{CA}$H and SiCNWN$_{CD}$H have more N dopants leading to a
higher carrier concentration of electrons and a larger chemical
potential. The ZT maximums are listed in the brackets in the figure
legends. It can be seen that for the n-type SiCNWN$_{C}$H, as the
chemical potential $\mu$ increases, the ZT values on the right side
of the highest peak are close to the ZT maximums of SiCNWN$_{CA}$H
and SiCNWN$_{CD}$H. Thus, as seen in Figs. 10(c) and 10(f), even if
the ZT curves of SiCNWN$_{CA}$H and SiCNWN$_{CD}$H are not shown,
for the cases with different doping concentrations of N, the
corresponding ZT could be approximately estimated from the ZT curve
of SiCNWN$_{C}$H.

From Figs. 10(c) and 10(f), it is clear that the ZT of 6 units is
larger than the ZT of 3 units for all the cases under interest. As
an example, the electronic and phonon transmission spectra of
SiCNWN$_{C}$V$^{Si}_{F}$H with different lengths have been
illustrated in Figs. 5(c) and 8. From the discussions in Figs. 3,
5(c), 6, 8 and 9, we know that for the cases of n-type doping, the
effect of strengthening the electron scattering is balanced by the
increase of the electron carriers for electronic transmission, while
the strengthening of the phonon scattering is unbalanced. Thus, as
the wire length increases from 2.3 to 4.6 nm, the electronic
transmission change little while the phonon transmission decreases
significantly, leading to the increase of ZT value.

In Sec. III, the localization length of the case with N dopant was
estimated to be greater than 100 nm at the bottom of conduction
band. Doping Si vacancies at the corners into the N doped SiCNW will
not change the localization length obviously, for
SiCNWN$_{C}$V$^{Si}_{F}$H has an electronic transmission spectrum
close to SiCNWN$_{C}$H. With a wire length smaller than the
localization length, the electronic resistance will increase along
with the wire length linearly. \cite{MFP1,MFP2} Moreover, it has
been reported in the case of SiNWs with a diameter of 1.2 nm that
the thermal conductance keeps decreasing dramatically before the
length increases over 50 nm, and after this length the thermal
conductance decreases almost linearly. \cite{SiNWte} In this sense,
for the nanowires such as SiCNWN$_{C}$V$^{Si}_{F}$H, the tendency of
ZT enhancement is probable to be kept not only for the wire length
from 2.3 nm to 4.6 nm, but also for a wire length of about 50 nm. As
a result, there is still a large room to expect a further
enhancement of ZT when the length increases from 4.6 nm, and a large
ZT can be maintained within the localization length.

\section{Summary}

In summary, the structural stability and the thermoelectric
properties of SiCNWs with N dopants and vacancies are investigated
by means of the density functional calculations. Our studies show
that the central (C) site for N dopants is energetically the most
stable. When the defects contain both N impurities and vacancies,
the most favorable configuration is that the N impurities are at the
center while the Si vacancies locate at the corners (F) with all the
dangling bonds passivated. Aiming at obtaining a large
thermoelectric figure of merit ZT, with N dopants at the center, a
small quantity of Si vacancies limited at the corners is most
favored for n-type, while a large quantity of Si vacancies spreading
to the flat edge (A) sites is most favored for p-type wires.

For the SiCNW as quantum wires along [111] with a diameter of 1.1 nm
and a length of 4.6 nm, the ZT maxima at 900 K are found to be about
1.78 for n-type and 1.66 for p-type, respectively. The ZT maximum of
n-type can be reached by the present n-type doping of N impurities
and Si vacancies, but for p-type it depends on whether or not there
are proper p-type dopants. As long as the decrease of phonon
transmission with the increase of length is far more rapid than the
decrease of electronic transmission, higher ZT values may be
expected for SiCNWs longer than 4.6 nm. Based on the reduced model
of quantum wires for simulation, the present findings may shed light
on the enhancement of the ZT values by taking advantage of N dopants
and surface Si vacancies in SiCNWs, that is especially favorable for
the thermoelectric performance at high temperature in applications.

\acknowledgments
We are grateful to Q. B. Yan, X. L. Sheng, X. Chen,
H. J. Cui and Eric Germaneau for useful discussions. All
calculations are completed on the supercomputer MagicCube
(DAWN5000A) in Shanghai Supercomputer Center. This work is supported
in part by the MOST, the NSFC (Grant Nos. 90922033, 10934008, and
10974253) and the CAS.

\end{document}